\documentclass[12pt,showkeys,amsmath,amssymb]{revtex4}
\usepackage{amsmath,amsfonts,amsthm,amscd,amssymb,latexsym}
\usepackage{bm}
\usepackage{dcolumn}
\usepackage{graphicx}
\usepackage{epstopdf}
\usepackage{color}
\usepackage{epsf}
\usepackage{epsfig}
\usepackage{graphicx, epic, eepic, color}
\usepackage[colorlinks,citecolor=blue,urlcolor=blue,linkcolor=blue]{hyperref}

\begin{document}

\title{G\"odel's Universe and the Local Limit of Nonlocal Gravity}

\author{Z. \surname{Mardaninezhad}$^1$}
\email{z.mardaninezhad@stud.uni-goettingen.de}
\author{B. \surname{Mashhoon}$^{2,3}$}
\email{mashhoonb@missouri.edu}

\affiliation{$^1$Faculty of Physics, Georg-August-University, D-37077 G\"ottingen, Germany \\
$^2$School of Astronomy, Institute for Research in Fundamental Sciences (IPM), Tehran 19395-5531, Iran\\
$^3$Department of Physics and Astronomy, University of Missouri, Columbia, Missouri 65211, USA\\
}

\date{\today}

\begin{abstract}
We consider the G\"odel universe within the context of the local limit of nonlocal gravity. This theory differs from Einstein's general relativity (GR) through the existence of a scalar susceptibility function $S(x)$ that is a characteristic feature of the gravitational field and vanishes in the GR limit. We show that G\"odel's spacetime is a solution of the modified gravitational field equations provided $S$ is a constant, in conformity with the spatial homogeneity of the G\"odel universe. 
\end{abstract}


\maketitle


\section{Introduction}

The purpose of this study is to determine if G\"odel's universe of general relativity~\cite{Einstein}  is also a solution of the gravitational field equations of the local limit of nonlocal gravity theory. The metric of G\"odel's spacetime is given by~\cite{Goedel}
\begin{equation}\label{I1} 
ds^2 = -\,dt^2 + 2\sqrt{2}\,(1-U)\,dt\,dy + dx^2 - (U^2 - 4 U + 2)\,dy^2 +dz^2\,,
\end{equation}
where 
\begin{equation}\label{I2} 
U(x) = e^{\sqrt{2}\,\Omega \,x}\,
\end{equation}
and $\Omega > 0$ is a constant parameter such that we recover the Minkowski spacetime for $\Omega = 0$.  The Cartesian system of coordinates $x^\mu = (t, x, y, z)$ in metric~\eqref{I1} is admissible. With $t \mapsto t- \sqrt{2}\,y$, the G\"odel metric assumes the simpler form
\begin{equation}\label{I3} 
ds^2 = -\,dt^2 -2\sqrt{2}\,U(x)\,dt\,dy + dx^2 -U^2(x)\,dy^2 +dz^2\,,
\end{equation}
which we henceforth employ throughout. Closed timelike curves do exist in the G\"odel universe~\cite{Goedel, HaEl, Stephani:2003tm, GrPo}. In this paper, we use natural units such that the speed of light in vacuum $c$ and Newton's constant of gravitation 
$G$ are set equal to unity, unless specified otherwise; moreover,  the metric signature is +2 and Greek indices run from 0 to 3, while Latin indices run from 1 to 3. 

The geometry of the G\"odel-type model has been studied by a number of authors~\cite{HaEl,  Stephani:2003tm, GrPo}. The Weyl curvature of G\"odel spacetime is of type D in the Petrov classification. The G\"odel universe admits five Killing vector fields, namely, $\partial_t$, $\partial_y$, $\partial_z$, $\partial_x - \sqrt{2}\,\Omega y\,\partial_y$ and~\cite{YNO, Chicone:2005vn} 
\begin{equation}\label{I4} 
 2\sqrt{2}\,U^{-1}\,\partial_t - 2\, \sqrt{2}\,\Omega y\,\partial_x + \left(2 \Omega^2y^2 - U^{-2}\right)\partial_y\,.  
\end{equation}

The G\"odel universe is a regular stationary and spatially homogeneous spacetime that contains rotating matter. Consider the class of observers that are all spatially at rest in this spacetime. Each such observer has a velocity 4-vector $u^\mu = \delta^\mu_0$ that is free of acceleration, expansion and shear; however, it is rotating in the negative sense about the $z$ axis and its vorticity 4-vector
\begin{equation}\label{I5} 
\omega^\mu  =  \frac{1}{2} \eta^{\mu \nu \rho \sigma}u_\nu  u_{\rho; \sigma}\, 
\end{equation}
is purely spatial; that is,  $\omega^\mu = (0, \bm{\omega})$ with the 3-vector $\bm{\omega}  =  (0, 0, -\,\Omega)$ in our Cartesian coordinate system. Here, $\eta_{\alpha \beta \gamma \delta} = (-g)^{\frac{1}{2}} \varepsilon_{\alpha \beta \gamma \delta}$ is the Levi-Civita tensor and  $\varepsilon_{\alpha \beta \gamma \delta}$ is the alternating symbol with $\varepsilon_{0123} = 1$ in our convention.

We are interested in the measurements of an observer that is free and spatially at rest in spacetime with 4-velocity vector $u^\mu = dx^\mu/d\tau$ and proper time $\tau$, where $\tau = t \,+\, {\rm constant}$. The observer carries along its geodesic world line a natural tetrad frame $e^{\mu}{}_{\hat \alpha}$ that is orthonormal, namely, 
\begin{equation}\label{I6} 
g_{\mu \nu}\, e^{\mu}{}_{\hat \alpha} \,e^{\nu}{}_{\hat \beta} = \eta_{\hat \alpha \hat \beta}\,, 
\end{equation}
where $\eta_{\mu \nu} = \rm{diag}(-1, 1, 1, 1)$ is the Minkowski metric tensor. Here, hatted indices specify the tetrad axes in the local tangent space; moreover, hatted indices are raised and lowered via $\eta_{\alpha \beta}$, while normal spacetime indices are raised and lowered via the spacetime metric $g_{\mu \nu}$.  Indeed, for the G\"odel metric~\eqref{I3},
\begin{equation}\label{I7} 
e_{\hat 0} = \partial_t\,, \qquad e_{\hat 1} = \partial_x\,, \qquad e_{\hat 2} =
-\,\sqrt{2}\,\partial_t + U^{-1}
\,\partial_y\,,\qquad e_{\hat 3} = \partial_z\,,
\end{equation}
where the spatial axes of the observer's frame are primarily along the background Cartesian coordinate axes. 

Let $\chi^{\mu}{}_{\hat \alpha}$ be the orthonormal tetrad frame that is parallel transported along the observer's geodesic world line such that $D\chi^{\mu}{}_{\hat \alpha}/d\tau = 0$. We find that 
\begin{equation}\label{I8} 
\chi^{\mu}{}_{\hat 1}  =  e^{\mu}{}_{\hat 1} \cos \Omega \tau + e^{\mu}{}_{\hat 2} \sin \Omega \tau\,, \qquad \chi^{\mu}{}_{\hat 2}  = - e^{\mu}{}_{\hat 1} \sin \Omega \tau + e^{\mu}{}_{\hat 2} \cos \Omega \tau\,,
\end{equation}
while  $\chi^{\mu}{}_{\hat 3} = e^{\mu}{}_{\hat 3}$ and naturally $\chi^{\mu}{}_{\hat 0} = e^{\mu}{}_{\hat 0} = u^\mu$. Therefore, the observer's natural frame rotates with respect to the parallel-transported frame about their common $z$ axis with frequency $-\Omega$, which is consistent with vorticity vector~\eqref{I5}. 

The field equations of GR are given by~\cite{Einstein}
\begin{equation}\label{I9} 
 {^0}G_{\mu \nu} + \Lambda g_{\mu \nu} = \kappa\,T_{\mu \nu}\,, \qquad \kappa:= \frac{8 \pi G}{c^4}\,,
\end{equation}
where ${^0}G_{\mu \nu}=~ ^0R_{\mu \nu} - \frac{1}{2} g_{\mu \nu}\, ^0R$ is the Einstein tensor and we use a left superscript ``0" to refer to geometric quantities directly derived from the Levi-Civita connection ($^0\Gamma^\mu_{\alpha \beta}$).  For the G\"odel universe, we have a comoving perfect fluid source
\begin{equation}\label{I10} 
T_{\mu \nu} = (\rho + p) u_\mu u_\nu + p\, g_{\mu \nu}\,, 
\end{equation}
where $\rho$ is the energy density, $p$ is the pressure and $u^\mu = \delta^\mu_0$ is the 4-velocity vector of the perfect fluid. In the case of G\"odel's universe, $^0R_{\mu \nu} = 2 \,\Omega^2 u_\mu u_\nu$ and 
\begin{equation}\label{I11} 
-\Lambda + \Omega^2  = \kappa\,\rho\,, \qquad  \Lambda + \Omega^2 =  \kappa\,  p\,.  
\end{equation}
In the absence of the cosmological constant $\Lambda$, we have as the source of  the G\"odel universe a perfect fluid with a stiff equation of state $\rho = p = \Omega^2/\kappa$. Another possibility is dust ($p = 0$), with $\tfrac{1}{2}\kappa \rho = -\Lambda = \Omega^2$. It follows from Eq.~\eqref{I11} that $-2\,\Lambda = \kappa (\rho - p)$; therefore, in any realistic situation ($\rho \ge p$), the cosmological constant of the G\"odel universe must be negative or zero ($\Lambda \le 0$).

We are interested in whether G\"odel's universe would fit into the framework of the local limit of nonlocal gravity. Nonlocal gravity (NLG) is a nonlocal generalization of GR patterned after the nonlocal electrodynamics of media~\cite{Hehl:2008eu, Hehl:2009es}. As is well known, it is possible to reformulate GR within the framework of teleparallelism~\cite{We, BlHe, Maluf:2013gaa, Aldrovandi:2013wha, Itin:2018dru}. The result is the teleparallel equivalent of GR (TEGR), which is the gauge theory of the 4-parameter Abelian group  of spacetime translations~\cite{Cho}. The TEGR field equations formally resemble Maxwell's equations of electrodynamics and can be made nonlocal in analogy with the electrodynamics of media~\cite{Hehl:2008eu, Hehl:2009es}. In this connection, let us recall that Maxwell's electrodynamics of media in an inertial frame of reference involves the electromagnetic field strength given by $(\mathbf{E}, \mathbf{B}) \mapsto F_{\mu\nu}$,
\begin{equation}\label{I12}
F_{\mu \nu} = \partial_{ \mu}A_{ \nu} -\partial_{ \nu}A_{ \mu}\,,
\end{equation}
where $A_\mu$ is the electromagnetic vector potential, as well as the electromagnetic field excitations $(\mathbf{D}, \mathbf{H}) \mapsto H_{\mu\nu}$ that include the  polarizability and magnetizability of the medium in response to $F_{\mu \nu}$. Indeed, 
\begin{equation}\label{I13}
\partial_{\nu}H^{\mu \nu} = \frac{4 \pi}{c} J^{\mu}\,,
\end{equation}
where $J^\mu$ is the conserved current 4-vector of \emph{free} electric charges in the inertial frame. Moreover, $F_{\mu \nu}$  and $H_{\mu\nu}$ are connected through the \emph{constitutive relations} that are characteristic of the electromagnetic properties of the material  medium. In general, the constitutive relations are nonlocal~\cite{Hop, Poi, Jack, L+L, HeOb, VanDenHoogen:2017nyy}. They assume the simple forms
\begin{equation}\label{I14}
\mathbf{D} = \epsilon(x)\, \mathbf{E}\,, \qquad  \mathbf{B} = \mu(x) \,\mathbf{H}\,
\end{equation}
in their local limits. Here,  $\epsilon$ and $\mu$ are the electric permittivity and magnetic permeability, respectively.

In the next section, we briefly outline the derivation of TEGR, its nonlocal extension, namely, nonlocal gravity (NLG), and its local limit. In the local limit of NLG, the modified Einstein field equations take on the form
\begin{equation}\label{I15} 
 {^0}G_{\mu \nu} + \Lambda g_{\mu \nu} - \kappa\,T_{\mu \nu} = \mathbb{M}_{\mu \nu}\,,
\end{equation}
where $\mathbb{M}_{\mu \nu}$ is, in general, not symmetric. In Section III, we work out the components of $\mathbb{M}_{\mu \nu}$ for the G\"odel universe. Section IV contains a brief discussion of our results. 


\section{Modified GR Framework}

Let us begin with GR's  tetrad description, which has a long history~\cite{BlHe, Maluf:2013gaa}. We imagine a congruence of observers in spacetime with adapted smooth tetrad frames $\lambda^\mu{}_{\hat \alpha}(x)$, where $x^\alpha = (t, x^i)$, $i = 1, 2, 3$, is an admissible coordinate system. For each observer, $\lambda^\mu{}_{\hat 0}(x)$ is the observer's unit temporal direction (i.e., 4-velocity vector) and $\lambda^\mu{}_{\hat i}(x)$, $i = 1, 2, 3$, form its spatial frame. In the tetrad theory, the metric is given by the orthonormality of the tetrad frame field, namely, 
\begin{equation}\label{M1}
g_{\mu \nu}(x) = \lambda_\mu{}^{\hat \alpha}(x)\, \lambda_\nu{}^{\hat \beta}(x)\, \eta_{\hat \alpha \hat \beta}\,.
\end{equation}

We can employ the tetrad frame field to define the Weitzenb\"ock connection~\cite{We}, 
\begin{equation}\label{M2}
\Gamma^\mu_{\alpha \beta}=\lambda^\mu{}_{\hat \rho }~\partial_\alpha\,\lambda_\beta{}^{\hat \rho}\,.
\end{equation}
The natural covariant derivative operator $\nabla$ associated with the Weitzenb\"ock connection has the property $\nabla_\nu\,\lambda_\mu{}^{\hat \alpha}=0$; that is, the tetrad frame field is covariantly constant. The curvature of the Weitzenb\"ock connection vanishes; therefore, the tetrad frames are parallel throughout the gravitational field. The resulting spacetime is a parallelizable manifold via the Weitzenb\"ock connection. Two distant vectors can be considered parallel if they have the same local components relative to the tetrad frame field. The Weitzenb\"ock connection thus gives rise to the framework of teleparallelism. Applying the $\nabla$ operator to the orthonormality relation~\eqref{M1}, we see that $\nabla_\mu\,g_{\alpha \beta} = 0$; that is, the Weitzenb\"ock connection is compatible with the spacetime metric $g_{\mu \nu}$. 

In the spacetime of modified GR under consideration here, free test particles and null rays follow geodesics of the spacetime manifold, namely, 
 \begin{equation}\label{M3}
\frac{d^2 x^\mu}{d\theta^2} +\, ^0\Gamma^\mu_{\alpha \beta}\, \frac{dx^\alpha}{d\theta}\,\frac{dx^\beta}{d\theta} = 0\,,
\end{equation}
where $\theta$ is the proper time (affine parameter) for a timelike (null) geodesic and 
\begin{equation}\label{M4}
^0\Gamma^\mu_{\alpha \beta} = \frac{1}{2} g^{\mu \nu}\,(g_{\nu \alpha, \beta} + g_{\nu \beta, \alpha} - g_{\alpha \beta, \nu})\,
\end{equation}
is the symmetric Levi-Civita connection. We thus have one metric and two metric-compatible connections in our spacetime. We can now define two new tensor fields, since the difference between two connections on the same manifold is always a tensor. In this way, we have the \emph{torsion} tensor
\begin{equation}\label{M5}
C_{\mu \nu}{}^{\rho}=\Gamma^{\rho}_{\mu \nu}-\Gamma^{\rho}_{\nu \mu}=\lambda^\rho{}_{\hat \beta }\left(\partial_{\mu}\lambda_{\nu}{}^{\hat \beta }-\partial_{\nu}\lambda_{\mu}{}^{\hat \beta}\right)\,
\end{equation}
and the \emph{contorsion} tensor 
\begin{equation}\label{M6}
K_{\mu \nu}{}^{\rho} =\, ^{0}\Gamma^{\rho}_{\mu \nu} - \Gamma^{\rho}_{\mu \nu}\,.
\end{equation}
The new tensors are related via $\nabla_\mu\,g_{\alpha \beta} = 0$, which implies $g_{\alpha \beta, \gamma} = \Gamma^{\mu}_{\gamma \alpha}\,g_{\mu \beta} + \Gamma^{\mu}_{\gamma \beta}\,g_{\mu \alpha}$. Using this relation in Eq.~\eqref{M4}, we find, with the help of Eqs.~\eqref{M5} and~\eqref{M6}, 
\begin{equation}\label{M7} 
K_{\mu \nu \rho} = \frac{1}{2}\, (C_{\mu \rho \nu}+C_{\nu \rho \mu}-C_{\mu \nu \rho})\,.
\end{equation}
Let us note that the torsion tensor $C_{\alpha \beta \gamma}$ is antisymmetric in its first two indices, $C_{\alpha \beta \gamma} = - C_{\beta \alpha \gamma}$, while the contorsion tensor $K_{\alpha \beta \gamma}$ is antisymmetric in its last two indices, 
$K_{\alpha \beta \gamma} = - K_{\alpha \gamma \beta}$.

\subsection{TEGR}

It follows from the definition of the contorsion tensor that the Christoffel symbols can be written in terms of the Weitzenb\"ock connection and torsion, since  $^{0}\Gamma^{\alpha}_{\mu \nu} = \Gamma^{\alpha}_{\mu \nu} + K_{\mu \nu}{}^{\alpha}$; therefore, it is possible to express the Riemann curvature tensor \,$^0R_{\alpha \beta \gamma \delta}$ in terms of the torsion tensor and associated quantities. Indeed, the Einstein tensor can be expressed in the form
\begin{align}\label{M8}
 {^0}G_{\mu \nu}=\frac{\kappa}{\sqrt{-g}}\Big[\lambda_\mu{}^{\hat{\gamma}}\,g_{\nu \alpha}\, \frac{\partial}{\partial x^\beta}\,\mathfrak{H}^{\alpha \beta}{}_{\hat{\gamma}}
-\Big(C_{\mu}{}^{\rho \sigma}\,\mathfrak{H}_{\nu \rho \sigma}
-\frac{1}{4}\,g_{\mu \nu}\,C^{\alpha \beta \gamma}\,\mathfrak{H}_{\alpha \beta \gamma}\Big) \Big]\,.
\end{align}
Here,  the auxiliary torsion field $\mathfrak{H}_{\mu \nu \rho}$ is defined via
\begin{equation}\label{M9}
\mathfrak{H}_{\mu \nu \rho}:= \frac{\sqrt{-g}}{\kappa}\,\mathfrak{C}_{\mu \nu \rho}\,, 
\end{equation}
where $\mathfrak{C}_{\alpha \beta \gamma}$ is the auxiliary torsion tensor  given by
\begin{equation}\label{M10}
\mathfrak{C}_{\alpha \beta \gamma} :=C_\alpha\, g_{\beta \gamma} - C_\beta \,g_{\alpha \gamma}+K_{\gamma \alpha \beta} = C_\alpha\, g_{\beta \gamma} - C_\beta \,g_{\alpha \gamma} +\frac{1}{2}\, (C_{\gamma \beta \alpha}+C_{\alpha \beta \gamma}-
C_{\gamma \alpha \beta})\,.
\end{equation}
In this relation, the torsion vector $C_\mu$ is defined by
\begin{equation}\label{M11}
C_\mu := C^{\alpha}{}_{\mu \alpha} = - C_{\mu}{}^{\alpha}{}_{\alpha}\,.
\end{equation}

Substituting the Einstein tensor~\eqref{M8} in the GR field Eq.~\eqref{I9}, we obtain the field equation of TEGR, namely, 
\begin{equation}\label{M12}
\frac{\partial}{\partial x^\nu}\,\mathfrak{H}^{\mu \nu}{}_{\hat{\alpha}}+\frac{\sqrt{-g}}{\kappa}\,\Lambda\,\lambda^\mu{}_{\hat{\alpha}} =\sqrt{-g}\,(T_{\hat{\alpha}}{}^\mu + \mathbb{T}_{\hat{\alpha}}{}^\mu)\,,
\end{equation}
where $\mathbb{T}_{\mu \nu}$ is the traceless energy-momentum tensor of the gravitational field given by
\begin{equation}\label{M13}
\mathbb{T}_{\mu \nu} := (\sqrt{-g})^{-1}\, (C_{\mu \rho \sigma}\, \mathfrak{H}_{\nu}{}^{\rho \sigma}-\tfrac{1}{4}  g_{\mu \nu}\,C_{\rho \sigma \delta}\,\mathfrak{H}^{\rho \sigma \delta})\,.
\end{equation}
The antisymmetry of $\mathfrak{H}^{\mu \nu}{}_{\hat{\alpha}}$ in its first two indices implies
\begin{equation}\label{M14}
\frac{\partial}{\partial x^\mu}\,\left[\sqrt{-g}\,(T_{\hat \alpha}{}^\mu - \tfrac{1}{\kappa}\,\Lambda\,\lambda^\mu{}_{\hat \alpha}+ \mathbb{T}_{\hat \alpha}{}^\mu)\right] = 0\,,
\end{equation}
which expresses the law of conservation of total energy-momentum tensor in TEGR. It seems that the problem of gravitational energy in GR has a solution in  TEGR~\cite{Maluf:2013gaa, Aldrovandi:2013wha, BlHe}. 

The procedure we have followed here that starts from GR and leads to TEGR would work for any smooth tetrad frame field $\lambda^\mu{}_{\hat \alpha}$. General relativity depends only upon the metric tensor $g_{\mu \nu}$ and is thus invariant under the \emph{local} Lorentz group. Similarly, TEGR, being essentially equivalent to GR, eventually depends only upon  $g_{\mu \nu}$; therefore, the tetrad formulation suffers from a 6-fold degeneracy (i.e., 3 boosts and 3 rotations of tetrads) at each spacetime event. 
For a detailed treatment, see~\cite{BMB} and the references therein. 

There is a formal analogy between TEGR and Maxwell's electrodynamics of media. Writing the torsion tensor $C_{\mu \nu}{}^{\hat \alpha} = \lambda_\rho{}^{\hat \alpha}\,C_{\mu \nu}{}^\rho$ via Eq.~\eqref{M5} as
\begin{equation}\label{M15}
C_{\mu \nu}{}^{\hat \alpha} = \partial_{\mu}\lambda_{\nu}{}^{\hat \alpha}-\partial_{\nu}\lambda_{\mu}{}^{\hat \alpha}\,,
\end{equation}
we note the similarity of $C_{\mu \nu}{}^{\hat \alpha}$, for each $\hat{\alpha}$, with $F_{\mu \nu}$ in Eq.~\eqref{I12}; moreover, this similarity extends to $\mathfrak{H}^{\mu \nu}{}_{\hat \alpha}$ and $H^{\mu \nu}$ in Eqs.~\eqref{M12} and~\eqref{I13}, respectively. From this standpoint, Eq.~\eqref{M9}, which connects $\mathfrak{H}_{\mu \nu \rho}$ with $C_{\mu \nu \rho}$, is the local constitutive relation of TEGR. 

As with the electromagnetic field tensor $F_{\mu \nu}$, the torsion field vanishes completely if the vector potential (i.e., $\lambda_{\mu}{}^{\hat \alpha}$) is a pure gauge, namely, $\lambda_{\mu}{}^{\hat \alpha}= \partial_\mu X^{\hat \alpha}$ for certain functions 
$X^{\hat \alpha}$. Then, the orthonormality condition~\eqref{M1} implies that we are in flat spacetime and hence  \,$^0R_{\alpha \beta \gamma \delta} = 0$. On the other hand, if \,$^0R_{\alpha \beta \gamma \delta} \ne 0$, the torsion tensor does not vanish. 
Therefore, the Riemann curvature tensor \,$^0R_{\alpha \beta \gamma \delta}$ of the symmetric Levi-Civita connection and the torsion tensor $C_{\alpha \beta \gamma}$ of the curvature-free Weitzenb\"ock connection turn out to be complementary aspects of the gravitational field in this modified GR framework~\cite{BMB}.

\subsection{Nonlocal Gravity (NLG)}

In Maxwell's electrodynamics of media, the constitutive relation is characteristic of the medium under consideration and is in general nonlocal. That is, in the description of the electrodynamics of different media, the field equations remain the same and only the constitutive relation changes. We adopt this procedure for the development of NLG~\cite{Hehl:2008eu, Hehl:2009es}. Therefore, we replace the auxiliary torsion field $\mathfrak{H}_{\mu \nu \rho}$ on both sides of Eq.~\eqref{M12} with 
$\mathcal{H}_{\mu \nu \rho}$ defined by
\begin{equation}\label{M16}
\mathcal{H}_{\mu \nu \rho} := \frac{\sqrt{-g}}{\kappa}(\mathfrak{C}_{\mu \nu \rho}+ N_{\mu \nu \rho})\,,
\end{equation}
where $N_{\mu \nu \rho} = - N_{\nu \mu \rho}$ is a tensor involving a certain causal average of the gravitational field over spacetime. Next, we hope that the existence of this nonlocality would remove the 6-fold degeneracy of TEGR and the solution of the field equations of NLG would result in an essentially unique tetrad frame field $e^{\mu}{}_{\hat \alpha}$ adapted to certain fundamental observers in spacetime. Henceforth, 
\begin{equation}\label{M17}
 \lambda^{\mu}{}_{\hat \alpha}|_{\rm TEGR} \to e^{\mu}{}_{\hat \alpha}|_{\rm NLG}\,,
\end{equation}
so that in NLG, we have the basic field equation 
\begin{equation}\label{M18}
 \frac{\partial}{\partial x^\nu}\,\mathcal{H}^{\mu \nu}{}_{\hat{\alpha}}+\frac{\sqrt{-g}}{\kappa}\,\Lambda\,e^\mu{}_{\hat{\alpha}} =\sqrt{-g}\,(T_{\hat{\alpha}}{}^\mu + \mathcal{T}_{\hat{\alpha}}{}^\mu)\,,
\end{equation}
where $\mathcal{T}_{\mu \nu}$ is now the traceless energy-momentum tensor of the gravitational field and involves the nonlocality tensor $N_{\mu \nu \rho}$. Specifically, 
\begin{equation}\label{M18a}
\mathcal{T}_{\mu \nu} := (\sqrt{-g})^{-1}\, (C_{\mu \rho \sigma}\, \mathcal{H}_{\nu}{}^{\rho \sigma}-\tfrac{1}{4}  g_{\mu \nu}\,C_{\rho \sigma \delta}\,\mathcal{H}^{\rho \sigma \delta})\,.
\end{equation}
The conservation law of the total energy-momentum is now given by
\begin{equation}\label{M19}
\frac{\partial}{\partial x^\mu}\,\Big[\sqrt{-g}\,(T_{\hat{\alpha}}{}^\mu + \mathcal{T}_{\hat{\alpha}}{}^\mu - \tfrac{1}{\kappa}\,\Lambda\,e^\mu{}_{\hat \alpha })\Big]=0\,.
 \end{equation}

It is useful to introduce a  traceless tensor $Q_{\mu \nu}$ such that
\begin{equation}\label{M20}
\mathcal{T}_{\mu \nu} - \mathbb{T}_{\mu \nu} = \frac{1}{\kappa}\,Q_{\mu \nu}\,,
\end{equation}
where $\mathbb{T}_{\mu \nu}$ is defined in Eq.~\eqref{M13} in terms of $\mathfrak{H}_{\mu \nu \rho}$ and
\begin{equation}\label{M21}
Q_{\mu \nu} := C_{\mu \rho \sigma} N_{\nu}{}^{\rho \sigma}-\frac 14\, g_{\mu \nu}\,C_{ \delta \rho \sigma}N^{\delta \rho \sigma}\,.
\end{equation} 
 
We must specify how the nonlocality tensor $N_{\mu \nu \rho}$ is related to the gravitational field. In NLG, we assume~\cite{Puetzfeld:2019wwo, Mashhoon:2022ynk} 
\begin{equation}\label{M22}
N_{\hat \mu \hat \nu \hat \rho}(x) = \int \mathcal{K}(x, x')\,\{\mathfrak{C}_{\hat \mu \hat \nu \hat \rho}(x')+ \check{p}\,[\check{C}_{\hat \mu}(x')\, \eta_{\hat \nu \hat \rho}-\check{C}_{\hat \nu}(x')\, \eta_{\hat \mu \hat \rho}]\} \sqrt{-g(x')}\, d^4x'\,,
\end{equation}
where $\mathcal{K}(x, x')$ is the causal kernel of the theory and $\check{p}\ne 0$ is a constant dimensionless parameter. Here, $\check{C}_\mu$ is the torsion pseudovector defined by 
\begin{equation}\label{M23}
\check{C}_\mu :=\frac{1}{3!} C^{\alpha \beta \gamma}\,\eta_{\alpha \beta \gamma \mu}\,,
\end{equation}
where $\eta_{\alpha \beta \gamma \delta}$ is the Levi-Civita tensor.  The causal kernel of NLG introduces the influence of the past history of the gravitational field (``memory") into the current field equations and thus renders them nonlocal. A comprehensive account of NLG is contained in~\cite{BMB}. 

Within the framework of TEGR, the gravitational field equations are equivalent to Einstein's field equations of GR. The nonlocal extension of TEGR, namely, NLG, should correspond to modified GR field equations. These can be determined by substituting for
$\mathfrak{H}_{\mu \nu \rho}$ in the Einstein tensor~\eqref{M8} the quantity $\mathcal{H}_{\mu \nu \rho} -\tfrac{\sqrt{-g}}{\kappa}\, N_{\mu \nu \rho}$ and using Eq.~\eqref{M18} to get
\begin{equation}\label{M24}
^{0}G_{\mu \nu} + \Lambda g_{\mu \nu} = \kappa T_{\mu \nu}   +  Q_{\mu \nu} -  \mathcal{N}_{\mu \nu}\,,
\end{equation}
where the new tensor $\mathcal{N}_{\mu \nu}$ is given by
\begin{equation}\label{M25}
\mathcal{N}_{\mu \nu} := g_{\nu \alpha} e_\mu{}^{\hat{\gamma}} \frac{1}{\sqrt{-g}} \frac{\partial}{\partial x^\beta}\,(\sqrt{-g}N^{\alpha \beta}{}_{\hat{\gamma}})\,.
\end{equation} 
The field equations of nonlocally modified GR are given by the addition of $Q_{\mu \nu} - \mathcal{N}_{\mu \nu}$ to the right-hand side of Einstein's field equations of GR. We need the 16 field equations~\eqref{M24} of NLG to determine the 16 components of the tetrad frame field $e^\mu{}_{\hat \alpha}$. Of the 16 tetrad components, ten determine the spacetime metric tensor $g_{\mu \nu}$ and the remaining 6 are local Lorentz  degrees of freedom (i.e., boosts and rotations); in fact, this division corresponds to splitting Eq.~\eqref{M24} into its symmetric and antisymmetric components. The 10 symmetric components of Eq.~\eqref{M24} in essence determine the metric tensor of NLG, while the 6 antisymmetric components $Q_{[\mu \nu]} = \mathcal{N}_{[\mu \nu]}$ are integral constraint equations. 

The field equations of NLG are highly nonlinear partial integro-differential equations  involving  an unknown kernel $\mathcal{K}(x, x')$. Therefore, it has not been possible to explore nonlinear regime of NLG involving black holes or cosmological models.  On the other hand,  linearized NLG  and its Newtonian regime have been the subject of extensive investigations.  At this stage of development of NLG, it appears that the kernel in the linear regime can be essentially determined via observation. In fact, in the linear regime of NLG and its Newtonian limit, the nonlocal aspect of gravity can simulate dark matter~\cite{Rahvar:2014yta, Chicone:2015coa, Roshan:2021ljs, Roshan:2022zov, Roshan:2022ypk}. Returning to the nonlinear regime, it  has not been possible to find exact nontrivial solutions of NLG. Indeed, Minkowski spacetime with $e^{\mu}{}_{\hat \alpha} = \delta^\mu_\alpha$ is the only known exact solution of NLG  in the absence of gravity and sources; moreover, de Sitter spacetime is not a solution of NLG~\cite{Mashhoon:2022ynk}.  

It is important to mention that there are indeed various other models of nonlocal gravity and cosmology; see, for instance~\cite{Maggiore:2014sia, Woodard:2018gfj, Deser:2019lmm, Balakin:2022gjw, Koshelev:2022bvg, Koshelev:2022olc, Jusufi:2023ayv, Bajardi:2024kea, Capozziello:2024qol} and the references cited therein. The classical approach adopted in the present work is uniquely based on the analogy with the nonlocal electrodynamics of media. 

In the electrodynamics of material media,  it is often necessary to employ  simple local relations $\mathbf{D} = \epsilon(x) \mathbf{E}$ and $\mathbf{B} = \mu (x) \mathbf{H}$ instead of the nonlocal constitutive relations. One expects that these local limits of nonlocal constitutive relations of material media express some important features of the general nonlocal situation. In a similar manner, we can resort to the local limit of NLG in order to explore its nonlinear regime. This is the approach that we adopt in the rest of this paper.

\subsection{Local Limit of NLG}

In the local limit of NLG, we assume that $\mathcal{K}(x, x')$ is proportional to the 4D Dirac delta function; that is, 
\begin{equation}\label{M26}
\mathcal{K}(x, x') := \frac{S(x)}{\sqrt{-g(x)}}\,\delta(x-x')\,,
\end{equation}
which defines a scalar \emph{susceptibility} function $S(x)$ that is characteristic of the background spacetime~\cite{Tabatabaei:2023qxw, Tabatabaei:2023lec, Tabatabaei:2022tbq, Tabatabaei:2023iwc}. We can compare the gravitational susceptibility  $S(x)$ to the electrical permittivity $\epsilon(x)$ and magnetic permeability $\mu(x)$ that are
 constitutive aspects of a medium in electrodynamics. The nonlocality tensor   $N_{\mu \nu \rho}$ thus reduces to a local tensor in this case,  
\begin{equation}\label{M27}
N_{\mu \nu \rho}(x) = S(x)\,[\mathfrak{C}_{\mu \nu \rho}(x) + \check{p}\,(\check{C}_\mu\, g_{\nu \rho}-\check{C}_\nu\, g_{\mu \rho})]\,
\end{equation}
and the local constitutive relation takes the form
\begin{equation}\label{M28}
\mathcal{H}_{\mu \nu \rho} = \frac{\sqrt{-g}}{\kappa}[(1+S)\,\mathfrak{C}_{\mu \nu \rho}+ S\,\check{p}\,(\check{C}_\mu\, g_{\nu \rho}-\check{C}_\nu\, g_{\mu \rho})]\,.
\end{equation}
For $S=0$, we recover the TEGR and hence GR; moreover, it follows from the comparison between Eqs.~\eqref{M28} and~\eqref{M9} that for $S \ne 0$ one can preserve the physical connection between the local limit of NLG and GR  if one assumes $1+S > 0$. 

What role does $S(x)$ play in gravitational physics? In this connection, the standard Friedmann-Lema\^itre-Robertson-Walker (FLRW) cosmological models have been studied within the framework of the local limit of NLG~\cite{Tabatabaei:2023qxw, Tabatabaei:2022tbq, Tabatabaei:2023iwc}. In particular, one finds that  the functions $S(t)$ and $dS(t)/dt$, where $t$ is cosmic time, appear in the modified expressions for the spatially uniform cosmic density and pressure of the flat benchmark model. These interesting alterations involve a dynamic dark energy component. Moreover, observational data have been employed in order to determine $S(t)$ in connection with the $H_0$ tension~\cite{Tabatabaei:2022tbq, Tabatabaei:2023iwc}. 
Similarly, in the case of the G\"odel universe, we would  expect changes in the expressions for the density and pressure given in Eq.~\eqref{I11}. 
 
The locally modified GR field equations for the local limit of NLG can be written in the same form as in Eq.~\eqref{M24}; however, in the present case the 16 components of 
\begin{equation}\label{M29}
Q_{\mu \nu} -  \mathcal{N}_{\mu \nu} = \mathbb{M}_{\mu \nu}\,,
\end{equation}
must be calculated using the locally defined $N_{\mu \nu \rho}$ given by Eq.~\eqref{M27}. Writing Eq.~\eqref{M29} in terms of its symmetric and antisymmetric components,  $\mathbb{M}_{\mu \nu} = \mathbb{M}_{(\mu \nu)} + \mathbb{M}_{[\mu \nu]}$, we note that $\mathbb{M}_{(\mu \nu)}$ would be involved in the symmetric part of the modified Einstein's field equations, while $\mathbb{M}_{[\mu \nu]} = 0$. We compute these components in the next section for the G\"odel universe.

\section{G\"odel's Universe and Extended GR}

G\"odel-type spacetimes have been discussed within the general context of teleparallel gravity by Obukhov and other investigators~\cite{Obukhov:2004hw, Ulhoa:2013hea, Sousa:2008av}. The man purpose of the present work is to find the conditions under which  G\"odel's solution of GR could also be a solution of the local limit of nonlocal gravity.  

For the calculations regarding the G\"odel universe, we collect here some useful relations:
\begin{equation}\label{T1} 
e^{\mu}{}_{\hat 0} = (1, 0, 0, 0)\,, \qquad  e^{\mu}{}_{\hat 1} = (0, 1, 0, 0)\,,
\end{equation}
\begin{equation}\label{T2} 
e^{\mu}{}_{\hat 2} = (-\sqrt{2}, 0, U^{-1}, 0)\,,\qquad  e^{\mu}{}_{\hat 3} = (0, 0, 0, 1)\,.
\end{equation}
Next, 
\begin{equation}\label{T3} 
e_{\mu}{}^{\hat 0} = (1, 0, \sqrt{2}\,U, 0)\,, \qquad  e_{\mu}{}^{\hat 1} = (0, 1, 0, 0)\,,
 \end{equation}
\begin{equation}\label{T4} 
e_{\mu}{}^{\hat 2} = (0, 0, U, 0)\,,\qquad  e_{\mu}{}^{\hat 3} = (0, 0, 0, 1)\,. 
\end{equation}
The frame field given by Eqs.~\eqref{T1}--\eqref{T4} is adapted to preferred comoving observers that are spatially at rest and employ natural local axes that point essentially along the coordinate directions. Is this the unique fundamental tetrad frame of the theory? To address the issue of uniqueness, a separate investigation is necessary that is beyond the scope of the present work. 

The  G\"odel metric in matrix form can be expressed as 
\begin{equation}\label{T5}
(g_{\mu \nu}) = 
\begin{bmatrix}
-1&0&-\sqrt{2} \,U&0 \\
0&1&0&0 \\
-\sqrt{2}\, U&0&-U^2&0 \\
0&0&0&1 
\end{bmatrix}
\,, \quad (g^{\mu \nu}) =
\begin{bmatrix}
1&0&-\sqrt{2}\,U^{-1}&0 \\
0&1&0&0 \\
-\sqrt{2}\, U^{-1}&0&U^{-2}&0 \\
0&0&0&1 
\end{bmatrix}\,,
\end{equation}
where $U(x) = e^{\sqrt{2}\Omega x}$ and $\sqrt{-g} = U(x)$.  

It is possible to show that for the G\"odel solution under consideration here the Weitzenb\"ock connection has only one nonzero component given by
\begin{equation}\label{T6} 
 \Gamma^2_{12} = \frac{U'}{U} = \sqrt{2}\,\Omega\,. 
\end{equation}
Henceforth a prime denotes differentiation with respect to the $x$ coordinate, i.e., $U' := dU/dx$. 

The nonzero components of the torsion tensor are then given by
\begin{equation}\label{T7} 
 C_{12}{}^2 = -C_{21}{}^{2} =  \sqrt{2}\,\Omega\,. 
\end{equation}
Hence, the only nonzero components of $C_{\mu \nu \rho}$ are
\begin{equation}\label{T8} 
 C_{120} = -C_{210} =  -2\,\Omega\,U\,, \qquad C_{122} = -C_{212} = - \sqrt{2}\,\Omega\,U^2\,. 
\end{equation}
The torsion vector $C_{\mu} = C^{\alpha}{}_{\mu \alpha}$ is thus given by
\begin{equation}\label{T9} 
 C_{\mu} = (0, -\sqrt{2}\,\Omega, 0, 0)\,, 
\end{equation}
while the torsion pseudovector can be evaluated using Eq.~\eqref{M23} and the result is 
\begin{equation}\label{T9a} 
 \check{C}_{\mu} = (0, 0, 0, \tfrac{2}{3}\,\Omega)\,. 
\end{equation}

The  nonzero components of the contorsion tensor~\eqref{M7} are given by
\begin{equation}\label{T10} 
 K_{012} = - K_{021} = -\Omega \,U\,, \qquad K_{120} = - K_{102} = \Omega\,U\,, 
\end{equation}
\begin{equation}\label{T11} 
 K_{210} = - K_{201} = -\Omega \,U\,, \qquad K_{212} = - K_{221} = -\sqrt{2}\,\Omega\,U^2\,. 
\end{equation}
Moreover, the nonzero components of the auxiliary torsion tensor~\eqref{M10} are given by 
\begin{equation}\label{T12} 
 \mathfrak{C}_{010} = - \mathfrak{C}_{100} = - \sqrt{2}\,\Omega\,, \qquad \mathfrak{C}_{120} = - \mathfrak{C}_{210} = \Omega\,U\,, 
\end{equation}
\begin{equation}\label{T13} 
 \mathfrak{C}_{021} = - \mathfrak{C}_{201} = -\Omega \,U\,, \qquad \mathfrak{C}_{012} = - \mathfrak{C}_{102} = -\Omega\,U\,, 
\end{equation}
\begin{equation}\label{T14} 
 \mathfrak{C}_{313} = - \mathfrak{C}_{133} = \sqrt{2}\,\Omega\,. 
\end{equation}

Let us define $\mathfrak{P}_{\mu \nu \rho}$,
\begin{equation}\label{T14a}
\mathfrak{P}_{\mu \nu \rho}(x) := \check{C}_\mu\, g_{\nu \rho}-\check{C}_\nu\, g_{\mu \rho}\,
\end{equation}
and note that Eq.~\eqref{M27} can now be written as
\begin{equation}\label{T14b}
N_{\mu \nu \rho} = S\,(\mathfrak{C}_{\mu \nu \rho} + \check{p}\,\mathfrak{P}_{\mu \nu \rho})\,,
\end{equation}
where $S$ can only depend on the $x$ coordinate in the case of G\"odel's spacetime. For the G\"odel solution, the nonzero components of $\mathfrak{P}_{\mu \nu \rho}$ are given by
\begin{equation}\label{T14c} 
 \mathfrak{P}_{030} = - \mathfrak{P}_{300} = \frac{2}{3}\,\Omega\,, \qquad \mathfrak{P}_{032} = - \mathfrak{P}_{302} = \frac{2\sqrt{2}}{3}\,\Omega\,U\,, 
\end{equation}
\begin{equation}\label{T14d} 
 \mathfrak{P}_{131} = - \mathfrak{P}_{311} = - \frac{2}{3}\Omega\,, \qquad \mathfrak{P}_{230} = - \mathfrak{P}_{320} = \frac{2\sqrt{2}}{3}\,\Omega\,U\,, 
\end{equation}
\begin{equation}\label{T14e} 
 \mathfrak{P}_{232} = - \mathfrak{P}_{322} = \frac{2}{3}\,\Omega\,U^2\,. 
\end{equation}

We are now ready to compute $Q_{\mu \nu} = \bar{Q}_{\mu \nu} + \check{Q}_{\mu \nu}$, where
\begin{equation}\label{T15}
\bar{Q}_{\mu \nu} := S\left(C_{\mu \rho \sigma} \mathfrak{C}_{\nu}{}^{\rho \sigma}-\tfrac{1}{4}\, g_{\mu \nu}\,C_{ \delta \rho \sigma}\mathfrak{C}^{\delta \rho \sigma}\right)\,,
\end{equation} 
\begin{equation}\label{T15a}
\check{Q}_{\mu \nu} := S\,\check{p}\,\left(C_{\mu \rho \sigma} \mathfrak{P}_{\nu}{}^{\rho \sigma}-\tfrac{1}{4}\, g_{\mu \nu}\,C_{ \delta \rho \sigma}\mathfrak{P}^{\delta \rho \sigma}\right)\,,
\end{equation} 
since $N_{\mu \nu \rho}$ in Eq.~\eqref{M21} is now given by Eq.~\eqref{T14b}. In connection with the calculation of $\bar{Q}_{\mu \nu}$, we have
\begin{equation}\label{T16}
\mathfrak{C}^{120} = \frac{\Omega}{U}\,, \qquad \mathfrak{C}^{122} = 0\,, \qquad C_{ \delta \rho \sigma}\mathfrak{C}^{\delta \rho \sigma} = -4 \Omega^2\,
\end{equation} 
and after some algebra, we eventually find
\begin{equation}\label{T17}
(\bar{Q}_{\mu \nu}) = S \Omega^2
\begin{bmatrix}
-1&0&-\sqrt{2} \,U&0 \\
0&-1&0&0 \\
0&0&-U^2&0 \\
0&0&0&1 
\end{bmatrix}
\end{equation}
and note that it is traceless, i.e. $g^{\mu \nu}\,\bar{Q}_{\mu \nu} = 0$. Turning our attention now to $\check{Q}_{\mu \nu}$, we note that the nonzero components of $\mathfrak{P}^{\mu \nu \rho}$ are given by
\begin{equation}\label{T17a} 
 \mathfrak{P}^{030} = - \mathfrak{P}^{300} = -\frac{2}{3}\,\Omega\,, \qquad \mathfrak{P}^{032} = - \mathfrak{P}^{302} = \frac{2\sqrt{2}}{3}\,\Omega\,U^{-1}\,, 
\end{equation}
\begin{equation}\label{T17b} 
 \mathfrak{P}^{131} = - \mathfrak{P}^{311} = - \frac{2}{3}\Omega\,, \qquad \mathfrak{P}^{230} = - \mathfrak{P}^{320} = \frac{2\sqrt{2}}{3}\,\Omega\,U^{-1}\,, 
\end{equation}
\begin{equation}\label{T17c} 
 \mathfrak{P}^{232} = - \mathfrak{P}^{322} = -\frac{2}{3}\,\Omega\,U^{-2}\,. 
\end{equation}
It follows from these results that $C_{ \delta \rho \sigma}\mathfrak{P}^{\delta \rho \sigma} = 0$; moreover,  $\check{Q}_{\mu \nu}$ has only one nonzero component, namely,  
\begin{equation}\label{T17d} 
\check{Q}_{13} = \frac{2\sqrt{2}}{3}\,\check{p}\,\Omega^2 S(x)\,. 
\end{equation}

Next, we need to compute 
\begin{equation}\label{T18}
\mathcal{N}_{\mu \nu} := g_{\nu \alpha} e_\mu{}^{\hat{\gamma}} \frac{1}{\sqrt{-g}} \frac{\partial}{\partial x^\beta}\,(\sqrt{-g}N^{\alpha \beta}{}_{\hat{\gamma}})\,,
\end{equation} 
where $\sqrt{-g} = U$ and $\beta = 1$, since the only dependence of $U$ and $S$ could be on the $x$ coordinate. That is, $S(x)$ is a characteristic of the G\"odel spacetime whose metric only depends on the $x$ coordinate; therefore, $S$ can only depend on $x$. Moreover, we note
\begin{equation}\label{T19}
N^{\alpha \beta}{}_{\hat{\gamma}} = S g^{\alpha \mu}\,g^{\beta \nu}(\mathfrak{C}_{\mu \nu \rho}+ \check{p}\,\mathfrak{P}_{\mu \nu \rho})\, e^{\rho}{}_{\hat \gamma}\,.
\end{equation} 
After some algebra, we finally find that the only nonzero elements of $\mathcal{N}_{\mu \nu}$ are
\begin{equation}\label{T20}
\mathcal{N}_{0 0} = - \sqrt{2}\,\Omega\, S'\,, \qquad \mathcal{N}_{0 2} = - \Omega\,U S'\,, \qquad \mathcal{N}_{2 0} = - \Omega\,U S' + \sqrt{2}\,\Omega^2 U S\,,
\end{equation} 
\begin{equation}\label{T21}
\mathcal{N}_{2 2} = - \sqrt{2}\,\Omega\,U^2 S' + 2\, \Omega^2 U^2 S\,, \qquad \mathcal{N}_{3 3} =  2 \Omega^2\, S + \sqrt{2}\,\Omega \, S'\,,
\end{equation} 
\begin{equation}\label{T21a}
\mathcal{N}_{1 3} = \frac{2}{3}\,\check{p}\,\Omega\,(S' + \sqrt{2}\,\Omega \, S)\,.
\end{equation} 

The field equations of the local limit of nonlocal gravity are the same as Einstein's equations except that the energy density and pressure could now depend upon $S$, $(\rho, p) \to (\rho_S, p_S)$, and on the right-hand side we have in addition $\mathbb{M}_{\mu \nu} = Q_{\mu \nu} - \mathcal{N}_{\mu \nu}$, which turns out to have the form of a symmetric matrix except for a term proportional to $\check{p}$, namely, 
\begin{equation}\label{T22}
(\mathbb{M}_{\mu \nu}) =  \Omega
\begin{bmatrix}
\sqrt{2}S'-\Omega S&0&U(S'-\sqrt{2}\Omega S)&0 \\
0&-\Omega S&0&-\tfrac{2}{3}\check{p}S' \\
U(S'-\sqrt{2}\Omega S)&0&U^2(\sqrt{2}S' - 3 \Omega S)&0 \\
0&0&0&-(\sqrt{2}S'+\Omega S) 
\end{bmatrix}\,.
\end{equation}

Finally, let us write Eq.~\eqref{I15} in the form
\begin{equation}\label{T23}
^0G_{\mu \nu} + \Lambda g_{\mu \nu} - \kappa T_{\mu \nu} = (2 \Omega^2-\kappa\, \rho_S - \kappa\, p_S )\,u_\mu u_\nu + (\Omega^2 +\Lambda -\kappa p_S )\,g_{\mu \nu} = \mathbb{M}_{\mu \nu}\,.
\end{equation} 
Since $u^\mu = \delta^\mu _0$, we find $u_\mu = (-1, 0, -\sqrt{2}\,U, 0)$. Writing out the components of Eq.~\eqref{T23}, we find
\begin{equation}\label{T24}
\Omega^2 - \Lambda - \kappa \rho_S  = \Omega\,(\sqrt{2}S'-\Omega S)\,,
\end{equation} 
\begin{equation}\label{T25}
\sqrt{2} U(\Omega^2 - \Lambda - \kappa \rho_S ) = \Omega\,U(S'-\sqrt{2}\Omega S)\,,
\end{equation} 
\begin{equation}\label{T26}
\Omega^2 + \Lambda - \kappa p_S  = -\Omega^2\,  S\,,\qquad \check{p}\,S' = 0\,,
\end{equation} 
\begin{equation}\label{T27}
\Omega^2 + \Lambda - \kappa p_S  = -\Omega\,(\sqrt{2}S'+ \Omega S)\,,
\end{equation}
\begin{equation}\label{T28}
3\,\Omega^2 - \Lambda - 2\,\kappa \rho_S  - \kappa \,p_S  = \Omega\,(\sqrt{2}S'- 3\,\Omega S)\,.
\end{equation}  

These equations are consistent provided 
\begin{equation}\label{T29}
S' = 0\,,
\end{equation} 
regardless of the magnitude of $\check{p}$. The constancy of the susceptibility function $S$ is in conformity with the spatial homogeneity of the G\"odel universe. Equation~\eqref{I11} must then be modified such that 
\begin{equation}\label{T30} 
-\Lambda + (1+S)\, \Omega^2  = \kappa\,\rho_S \,, \qquad  \Lambda +(1+S)\, \Omega^2 =  \kappa\,  p_S \,.  
\end{equation}

The argument presented in~\cite{Mashhoon:2022ynk} shows specifically that de Sitter spacetime (with positive cosmological constant) is not a solution of nonlocal gravity. It is interesting to note that the cosmological constant that appears in Eq.~\eqref{T30} is such that $-2 \Lambda = \kappa (\rho_S - p_S)$ and, as discussed before, is expected in physically reasonable circumstances to be negative (or zero) and hence would correspond to the anti-de Sitter spacetime.  


\section{DISCUSSION}

Nonlocal gravity (NLG) is a classical extension of general relativity patterned after the electrodynamics of nonlocal media.  The local limit of NLG is a tetrad theory with 16 gravitational field equations for the 16 components of the fundamental tetrad frame field $e^\mu{}_{\hat \alpha}$.  This modified theory differs from general relativity due to the existence of a scalar gravitational susceptibility function $S(x)$ that is a characteristic feature of the spacetime in this theory. The situation here is analogous to employing local electric permittivity $\epsilon(x)$ and magnetic permeability $\mu(x)$ in the description of electromagnetic media.  We show that the G\"odel universe can be a solution of this theory provided $S$ is a constant, which is consistent with the temporal stationarity and spatial homogeneity of the G\"odel universe. With $S \ne 0$, the results are physically sensible for $1+S > 0$. In the G\"odel  spacetime of the local limit of NLG, we show that energy density and pressure are modified in accordance with 
$\Omega \mapsto (1+S)^{\frac{1}{2}}\,\Omega$, where $\Omega$ is the uniform angular speed of rotation of the G\"odel universe.


 

\appendix



\end{document}